\documentclass[preprint,trackchanges]{aastex7}

\accepted{September 23, 2025}

\begin{document}

\title{Variations of the vector magnetic structures in the solar polar regions observed by Hinode}

\author{Shuhong Yang}
\affiliation{State Key Laboratory of Solar Activity and Space Weather, National Astronomical Observatories, Chinese Academy of Sciences, Beijing 100101, China}
\affiliation{School of Astronomy and Space Science, University of Chinese Academy of Sciences, Beijing 100049, China}
\email[show]{shuhongyang@nao.cas.cn}

\author{Chunlan Jin}
\affiliation{State Key Laboratory of Solar Activity and Space Weather, National Astronomical Observatories, Chinese Academy of Sciences, Beijing 100101, China}
\affiliation{School of Astronomy and Space Science, University of Chinese Academy of Sciences, Beijing 100049, China}
\email{cljin@nao.cas.cn}

\author{Qiao Song}
\affiliation{Key Laboratory of Space Weather, National Satellite Meteorological Center (National Center for Space Weather), China Meteorological Administration, Beijing 100081, China}
\affiliation{Innovation Center for FengYun Meteorological Satellite (FYSIC), Beijing 100081, China}
\email{songq@cma.gov.cn}

\author{Yuzong Zhang}
\affiliation{State Key Laboratory of Solar Activity and Space Weather, National Astronomical Observatories, Chinese Academy of Sciences, Beijing 100101, China}
\affiliation{School of Astronomy and Space Science, University of Chinese Academy of Sciences, Beijing 100049, China}
\email{yuzong@nao.cas.cn}

\author{Yin Li}
\affiliation{State Key Laboratory of Solar Activity and Space Weather, National Astronomical Observatories, Chinese Academy of Sciences, Beijing 100101, China}
\affiliation{School of Astronomy and Space Science, University of Chinese Academy of Sciences, Beijing 100049, China}
\email{liyin277130@163.com}

\author{Yijun Hou}
\affiliation{State Key Laboratory of Solar Activity and Space Weather, National Astronomical Observatories, Chinese Academy of Sciences, Beijing 100101, China}
\affiliation{School of Astronomy and Space Science, University of Chinese Academy of Sciences, Beijing 100049, China}
\email{yijunhou@nao.cas.cn}

\author{Ting Li}
\affiliation{State Key Laboratory of Solar Activity and Space Weather, National Astronomical Observatories, Chinese Academy of Sciences, Beijing 100101, China}
\affiliation{School of Astronomy and Space Science, University of Chinese Academy of Sciences, Beijing 100049, China}
\email{liting@nao.cas.cn}

\author{Guiping Zhou}
\affiliation{State Key Laboratory of Solar Activity and Space Weather, National Astronomical Observatories, Chinese Academy of Sciences, Beijing 100101, China}
\affiliation{School of Astronomy and Space Science, University of Chinese Academy of Sciences, Beijing 100049, China}
\email{gpzhou@nao.cas.cn}

\author{Yuanyong Deng}
\affiliation{State Key Laboratory of Solar Activity and Space Weather, National Astronomical Observatories, Chinese Academy of Sciences, Beijing 100101, China}
\affiliation{School of Astronomy and Space Science, University of Chinese Academy of Sciences, Beijing 100049, China}
\email{dyy@nao.cas.cn}

\author{Jingxiu Wang}
\affiliation{State Key Laboratory of Solar Activity and Space Weather, National Astronomical Observatories, Chinese Academy of Sciences, Beijing 100101, China}
\affiliation{School of Astronomy and Space Science, University of Chinese Academy of Sciences, Beijing 100049, China}
\email{wangjx@nao.cas.cn}

\begin{abstract}

Using the polar vector magnetic field data observed by Hinode from 2012 to 2021, we study the long-term variations of the magnetic flux, the flux proportion of different polarities, and the magnetic inclination with respect to the local normal in the solar polar regions above 70$\degr$ latitude during solar cycle 24.
In both polar regions after the polarity reversal, the unsigned magnetic fluxes of the dominant polarity increased to a peak of about 1.3$\times$10$^{22}$ Mx during the solar minimum, while those of the non-dominant polarity remained stable at approximately 0.5$\times$10$^{22}$ Mx.
The proportions of the dominant polarity flux in the total flux in both polar regions increased to more than 70\% at the solar minimum.
These results reveal that the total magnetic flux and the number of open field lines in solar cycle 24 were larger than those in solar cycle 23, and also imply the existence of a local dynamo in polar regions.
After the polarity reversal, the magnetic inclination of the dominant polarity fields decreased, indicating that the stronger the dominant polarity field, the more vertical the field lines.
The inclination angle decreased with the increase of the threshold of radial magnetic flux density, revealing a fanning-out structure of the polar magnetic patches.

\end{abstract}



\section{Introduction}

The Sun's polar regions are considered crucial for understanding deep convection dynamics \citep{2023BAAS...55c.105F} and provide seed magnetic fields for the global dynamo that generates the solar cycle \citep{2020LRSP...17....4C,2020LRSP...17....2P}.
During solar minimum, each solar pole is covered by a huge coronal hole where the plasma escapes rapidly into interplanetary space along open magnetic field lines, forming a continuous and stable high-speed solar wind \citep{2005Sci...308..519T,2009LRSP....6....3C}. Studying the polar magnetic fields is important for revealing the origin of the solar magnetic cycle, clarifying the acceleration mechanisms of the high-speed solar wind, and understanding the structure of the heliosphere \citep{2023ChSBu...68..298D}. Because current solar polar observations are limited to the perspective from the ecliptic plane, many properties of the polar magnetic field remain unclear \citep{2015LRSP...12....5P,2023BAAS...55c.287N}.

Vector magnetic field observations provide detailed information on magnetic structures.
For the first time, \cite{1999ScChA..42.1096D} systematically measured the polar vector magnetic field on 1997 April 12 using the ground-based telescope at the Huairou Solar Observing Station. They found that the magnetic inclination of strong polar magnetic elements decreases with the increase of magnetic flux density.
With the polar vector magnetic field data from Hinode \citep{2007SoPh..243....3K}, \cite{2008ApJ...688.1374T} found that the strong magnetic elements in the south polar region exhibit small magnetic inclination angles relative to the local normal direction.
\cite{2010ApJ...719..131I} analyzed the polar magnetic field data observed by Hinode on 2007 September 25. Using the potential field source surface model to reconstruct the coronal magnetic field in the north polar region, they found that the majority of field lines from the kilogauss magnetic patches are open with fanning-out structures in the lower atmosphere.
Using the full spectropolarimetric data obtained by the Vacuum Tower Telescope in 2013, \cite{2018A&A...616A..46P} studied the polar magnetic topology near an activity maximum. In addition to some polar magnetic fields characterized by strong vertical fields, a new population of magnetic fields which were compatible with the presence of unresolved small-scale magnetic loops close to the limb were also inferred. Furthermore, \cite{2020A&A...635A.210P} investigated the magnetic topology of the north polar region in 2015, close to a maximum of activity. They found that most of the studied magnetic structures have a vertical flux of between 10$^{17}$ Mx and 10$^{19}$ Mx. With the high spatial resolution observations from the balloon-borne SUNRISE observatory, \cite{2020A&A...644A..86P} studied the fine structures of the polar magnetic fields, and compared them with the Hinode observations. They inferred kilogauss magnetic patches harboring polar faculae without applying a filling factor, and found that the vertical fields dominated the strong magnetic field regime. The horizontal fields were found to be ubiquitous for magnetic field strengths below about 300 G.

The polar magnetic fields at the solar minimum are considered to be the seed magnetic fields for the global dynamo process that generates the solar cycle \citep{2007MNRAS.381.1527J,2016ApJ...823L..22C}. At the solar minimum, the magnetic fields in each polar region are predominantly unipolar and reach their maximum strength \citep{2019SoPh..294..137P}.
\cite{1999ScChA..42.1096D} analyzed the polar magnetic field observations at the minimum phase of solar cycle 22 and found that the unsigned and net magnetic fluxes within the south polar cap above the latitude of 50$\degr$ were about 5.5$\times$10$^{22}$ Mx and $-$2.5$\times$10$^{22}$ Mx, respectively.
At the minimum of solar cycle 23, \cite{2008ApJ...688.1374T} investigated the vector magnetic field data of the south polar region obtained by Hinode on 2007 March 16, and found that many magnetic elements with strengths as high as 1 kilogauss are distributed within the latitude range of 70$-$90$\degr$. Their results also revealed that the total unsigned radial magnetic flux in the south polar region in the 70$-$90$\degr$ latitude range scarcely corresponded to the total magnetic flux of a single active region.
\cite{2010ApJ...719..131I} studied the north polar magnetic fields with Hinode observations on 2007 September 25, and found that, in the field-of-view corresponding to the surface area of 8.4$\times$10$^{20}$ cm$^{2}$, the net magnetic flux was 1.7$\times$10$^{21}$ Mx.

During the most time of solar cycle except for the period around the polar magnetic polarity reversal, the magnetic fields in both the north and south polar regions are dominated by one polarity \citep{2015LRSP...12....5P}.
With the Vacuum Tower Telescope observations in 2005, \cite{2007A&A...474..251B} investigated polar faculae that represent large-scale unipolar strands in the polar regions. Most of the polar faculae had the same polarity as the global magnetic field, and their sizes were larger than those with polarity opposite to the global field. \cite{2010A&A...509A..92B} studied the magnetic field in polar faculae, and found two distributions of the polar magnetic field structures, peaking at 400$-$600 G and at approximately 1200 G. Their results also revealed that a large fraction (more than 85\%) of polar faculae with strong fields have the same polarity as the global magnetic field around the poles.
Using the magnetic field data from the Solar and Heliospheric Observatory, \cite{2009SoPh..260..289L} found that during the minimum of solar cycle 23, the ratio of the number of magnetic elements with dominant polarity to that with opposite polarity in the polar regions was 3:1.
\cite{2011ApJ...732....4J} investigated the magnetic field data of the north polar coronal hole region observed by Hinode on 2007 September 10, and found that the ratio of magnetic flux between the dominant polarity and the minority polarity was approximately 2:1. \cite{2012ApJ...753..157S} studied the polar magnetic fields observed by Hinode from 2008 to 2012, and found that almost all large patches ($\ge$ 10$^{18}$ Mx) had the same polarity, while smaller patches had a fair balance of both polarities. During the solar maximum, when the polar magnetic fields are about to reverse, the numbers of the positive and negative magnetic field elements are almost balanced \citep{2009SoPh..260..289L}.

Hinode has routinely provided vector magnetic field observations of the solar polar regions over a solar cycle. These high-quality polar observations provide a unique opportunity to study the long-term variations of the polar vector magnetic structures, especially the inclination angles with respect to the local normal. Moreover, the long-term variations of the magnetic flux and the flux proportion of different polarities can be quantitatively determined. The present paper aims to investigate these aspects using the Hinode polar observations from 2012 to 2021.

\section{Observations and Data Analysis}

\begin{figure*}
\centering
\includegraphics
[width=0.75\textwidth]{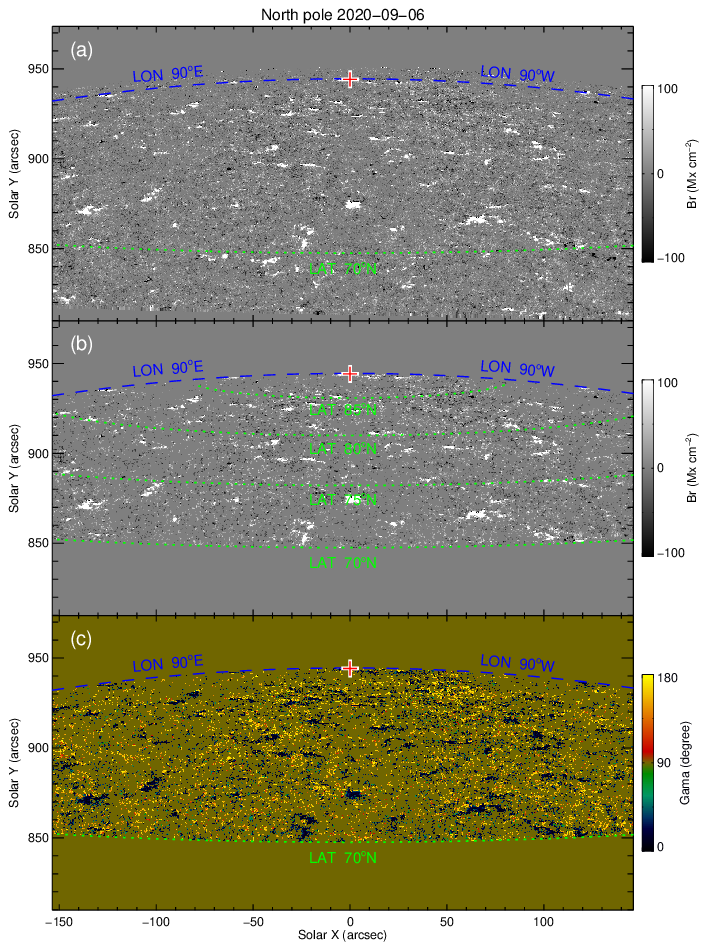} \caption{Overview of the north polar region at the solar minimum between solar cycles 24 and 25. (\textbf{a}) Radial magnetic flux density observed by Hinode/SP on 2020 September 6. The plus sign marks the pole. The green dotted curve indicates the 70$\degr$ latitude, the blue dashed curve outlines the boundary of $\pm$90$\degr$ longitude, and the region outside this range is not considered. (\textbf{b}) Magnetic patches with magnetic flux density stronger than 10 Mx cm$^{-2}$ and a size of no less than 3 contiguous pixels. (\textbf{c}) Magnetic inclination angles with respect to the local normal corresponding to the magnetic patches in (\textbf{b}). \label{fig_overview}}
\end{figure*}

\begin{figure*}
\centering
\includegraphics
[width=0.62\textwidth]{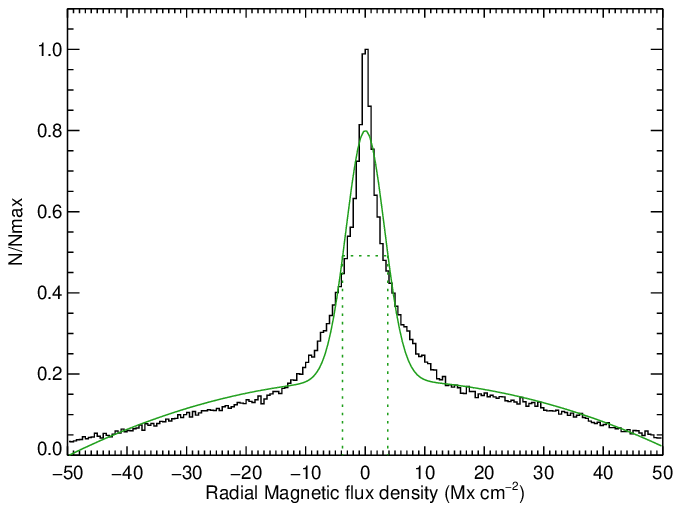} \caption{Distribution function of the polar radial magnetic fields observed by Hinode/SP. The green solid curve shows the Gaussian fit to the distribution function in the range of $\pm$50 Mx cm$^{-2}$. The horizontal dotted line marks the FWHM, i.e., 2$\sigma$. \label{fig_noise}}
\end{figure*}

As one of the most important instruments on board Hinode, the Spectro-polarimeter (SP; \citealt{2013SoPh..283..579L}) of the Solar Optical Telescope \citep{2008SoPh..249..167T} measures the photospheric vector magnetic fields with high polarimetric precision and spatial resolution. In this study, we adopt the Hinode/SP polar vector magnetic field data released at Nagoya University\footnote[1]{\url{https://hinode.isee.nagoya-u.ac.jp/sot\_polar\_field/}}.
With the Milne-Eddington Inversion of Polarized Spectra code \citep{2007A&A...462.1137O}, the full stokes profiles of Fe~{\sc{i}} 630.15 nm and 630.25 nm obtained by Hinode/SP were inverted to retrieve the magnetic field vector and the corresponding filling factor by applying a least-square fitting. For the transverse magnetic fields, the 180$\degr$ ambiguity was resolved with the method described in \cite{2010ApJ...719..131I} and \cite{2012ApJ...753..157S}.
One assumption is introduced in the disambiguation method, i.e., the magnetic field is either ``vertical'' or ``horizontal'' to the local surface (or undetermined), following \cite{2007ApJ...670L..61O} and \cite{2009A&A...495..607I}. For the vertical field, the zenith (inclination) angle is defined to be from 0$\degr$ to 40$\degr$ and from 140$\degr$ to 180$\degr$. For the horizontal field, the zenith angle is defined to be between 70$\degr$ and 110$\degr$. Due to the 180$\degr$ ambiguity of the transverse magnetic field, there are two solutions for the zenith angle in each pixel. According to \cite{2010ApJ...719..131I}: ``If both the solutions are vertical, the one closer to the local normal is taken. In case one solution is vertical and the other horizontal, it is not possible to distinguish one from each other. Pixels with those solutions are not used. If one of the solutions is either vertical or horizontal and the other solution is neither vertical nor horizontal, we will choose the solution of either vertical or horizontal.''

The vector magnetic field data of the south and north polar regions were mainly obtained in March and September, respectively. The SP slit scanned the polar regions in the east-west direction with a scanning step of about 0.30{\arcsec}, and the pixel size along the slit is approximately 0.32{\arcsec}.
The field of view is $\sim$ 300{\arcsec} (east-west along the scanning direction) $\times$ 163.84{\arcsec} (south-north along the slit).
The datasets of the polar magnetic field observations adopted here are the same as those used in \cite{2024RAA....24g5015Y,2024ApJ...970..183Y}. For each polar region every year, a dataset consisting of about 10 magnetograms is used \citep[see Table 1 in][]{2024RAA....24g5015Y}.
In order to calculate magnetic flux, the pixel size has been corrected for foreshortening effect with being divided by $\cos(\alpha)$, where $\alpha$ is the heliocentric angle of the pixel.

During the minimum of solar cycle 24, the magnetic fields in the north and south polar regions are predominated by the positive and negative polarities, respectively.
The overview of the radial magnetic fields in the north polar region observed by Hinode/SP on 2020 September 6 is shown in Figure \ref{fig_overview}a. The magnetogram shows that almost all the large magnetic patches in the north polar region have the positive polarity. In contrast, for the small magnetic patches, both the positive and negative polarities are presented. Due to the existence of the angle between the solar rotation axis and the normal to the ecliptic plane, the north pole (marked with the plus sign) is located within the visible solar disk in September. Because the magnetic field measurements for areas near the solar limb are not reliable, we do not consider regions beyond $\pm$90$\degr$ longitude (indicated by the blue dashed curve). In addition, the polar regions with latitudes lower than 70$\degr$ are also excluded.

In order to determine the noise level of the polar magnetic fields, we analyze the distribution function of the radial magnetic fields (see Figure \ref{fig_noise}). Considering the shape of the magnetic field distribution (see the black curve), we fit the observed magnetic data in the range of $\pm$50 Mx cm$^{-2}$ with a Gaussian function (indicated by the green solid curve), and the full width at half maximum (FWHM) is found to be 7.6 Mx cm$^{-2}$. If 1$\sigma$ is defined as FWHM/2, then $\sigma$ = 3.8 Mx cm$^{-2}$. In this study, 10 Mx cm$^{-2}$ (about 2.6$\sigma$) is used as the noise level.
Since the local magnetic inclination depends on the position over the solar disk, the line-of-sight uncertainties are propagated non-linearly and with position dependency into radial magnetic field. Moreover, the Hinode/SP noise level changes with time \citep{2012ApJ...753..157S}. Thus, we note that using a unique noise level implies considering noisier data close to the limb that farther away from it.

In the polar regions, we only consider the pixels with radial flux density stronger than 10 Mx cm$^{-2}$. Furthermore, the magnetic patches with fewer than 3 contiguous pixels are excluded, as shown in Figure \ref{fig_overview}b. 22\% of the pixels which pass all the thresholds are used. It is evident that both the positive and negative magnetic patches with the field strength exceeding the noise threshold are widespread. The magnetic inclination angles with respect to the local normal are reconstructed from the magnetic vectors, as shown in Figure \ref{fig_overview}c.
For strong magnetic elements (appearing as large white patches in Figure \ref{fig_overview}b), their inclination angles are relatively small (see the black patches in Figure \ref{fig_overview}c).

For the Sun with a radius of 6.96$\times$10$^{10}$ cm, the surface areas in each 5$\degr$ from 70$\degr$ latitude to the pole, i.e., 70$-$75$\degr$, 75$-$80$\degr$, 80$-$85$\degr$, and 85$-$90$\degr$, are 7.98$\times$10$^{20}$, 5.75$\times$10$^{20}$, 3.47$\times$10$^{20}$, and 1.16$\times$10$^{20}$ cm$^{2}$, respectively.
In a given magnetogram (as illustrated by Figure \ref{fig_overview}b in this study or Figure 1 in \citealt{2024ApJ...970..183Y}), the observed areas after foreshortening correction in these four latitude ranges are only about 1.36$\times$10$^{20}$, 1.50$\times$10$^{20}$, 1.63$\times$10$^{20}$, and 0.58$\times$10$^{20}$ cm$^{2}$, i.e., 0.17, 0.26, 0.47, and 0.5 times the corresponding solar surface areas, respectively. In addition, as revealed in \cite{2024ApJ...970..183Y}, the magnetic flux density in solar polar caps decreases from the lower latitudes to the poles. These two factors require us to apply an area-weighted metric in calculating the magnetic parameters in the entire north and south polar caps.
For a batch of \emph{n} (around 10 here) frames of magnetograms, the ratios of the total observed areas to surface areas for the four latitude ranges are \emph{n} $\times$ (0.17, 0.26, 0.47, and 0.5), respectively.
The magnetic parameters in each latitude range are calculated using all the \emph{n} frames together and then normalized to the surface area of the corresponding latitude range. For each batch of magnetograms, the magnetic parameters within the four latitude ranges are combined with the area-weighted metric to produce the results presented in this study.

\section{Results}

\begin{figure*}
\centering
\includegraphics
[width=0.67\textwidth]{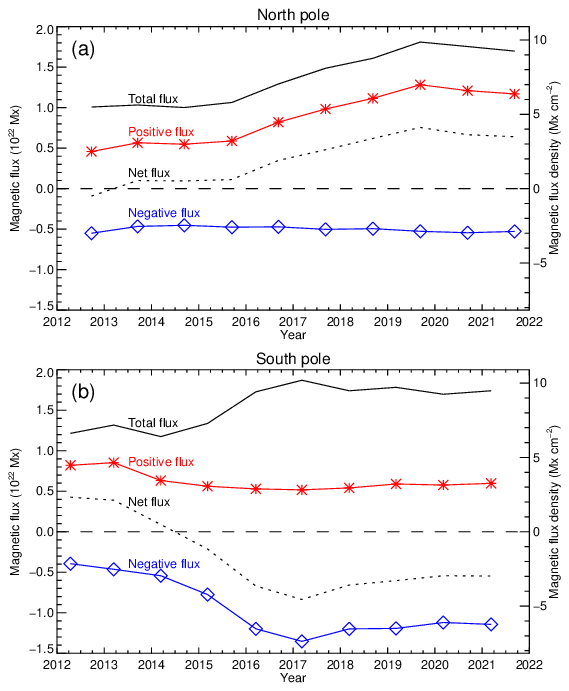} \caption{Variations of the magnetic flux and magnetic flux density in the polar regions above 70$\degr$ latitude from 2012 to 2021. (\textbf{a}) Positive flux (indicated by asterisks), negative flux (indicated by diamonds), total flux (black solid curve), and net flux (dotted curve) in the north polar region. (\textbf{b}) Similar to (\textbf{a}) but for the south polar region. \label{fig_flux_variation}}
\end{figure*}

\begin{figure*}
\centering
\includegraphics
[width=\textwidth]{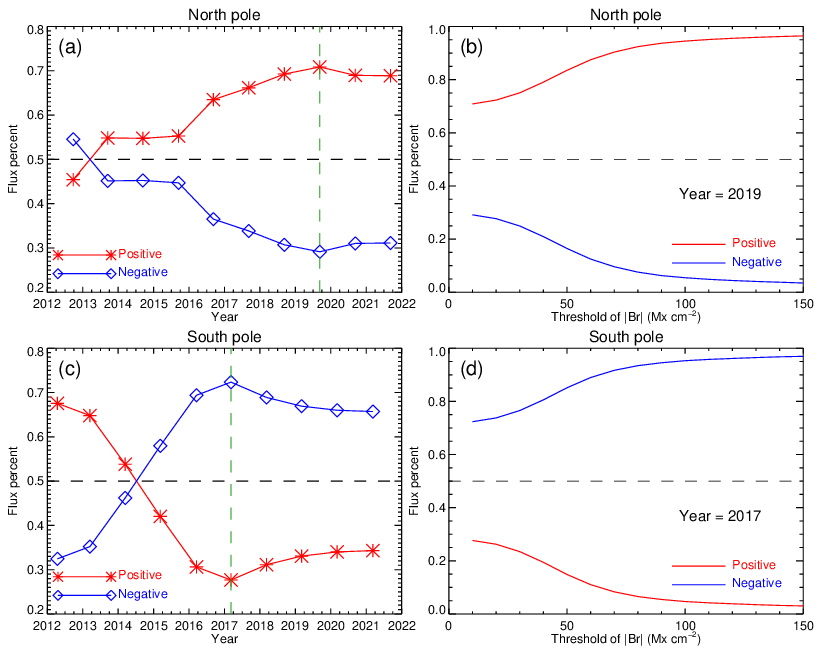} \caption{Proportions of the positive and negative magnetic flux in the total flux in polar regions.
(\textbf{a}) Long-term variations of the proportion of the positive flux (indicated by asterisks) and negative flux (indicated by diamonds) in the total flux in the north polar region. (\textbf{b}) Variations of the proportion of the positive flux (red curve) and negative flux (blue curve) in the total flux versus the thresholds of unsigned radial magnetic flux density in the north polar region in 2019 (marked by the vertical line in (\textbf{a})). (\textbf{c}) Similar to (\textbf{a}), but for the flux proportions in the south polar region. (\textbf{d}) Similar to (\textbf{b}), but for the south polar region in 2017 (marked by the vertical line in (\textbf{c})). The horizontal lines in (\textbf{a}) and (\textbf{c}) indicate the values at which the positive flux and negative flux are balanced. \label{fig_flux_percent}}
\end{figure*}

\begin{figure*}
\centering
\includegraphics
[width=\textwidth]{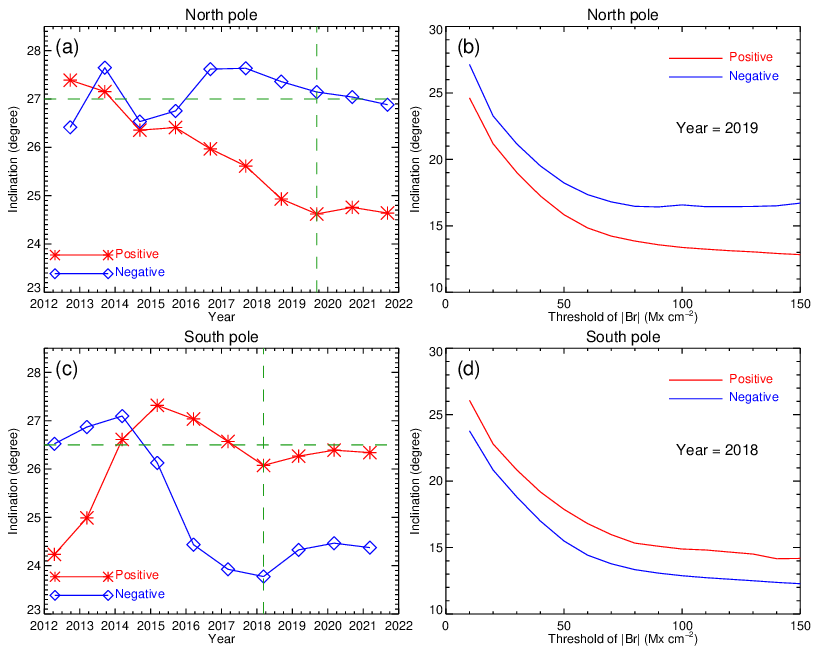} \caption{Magnetic inclination with respect to the local normal. (\textbf{a}) Long-term variations of the average inclination angles of the positive (indicated by asterisks) and negative (indicated by diamonds) magnetic fields in the north polar region. (\textbf{b}) Variations of the inclination angles of the positive (red curve) and negative (blue curve) fields versus the thresholds of unsigned radial flux density in the north polar region in 2019 (marked by the vertical line in (\textbf{a})). (\textbf{c}) Similar to (\textbf{a}), but for the magnetic inclination in the south polar region. (\textbf{d}) Similar to (\textbf{b}), but for the south polar region in 2018 (marked by the vertical line in (\textbf{c})). In this figure, the values of magnetic inclination of the negative fields are shown as 180$\degr$$-$$\gamma$, where $\gamma$ is the inclination angle with respect to the local normal. The horizontal lines in (\textbf{a}) and (\textbf{c}) indicate the values around which the inclination angles of non-dominant polarity fluctuate. \label{fig_inclination}}
\end{figure*}

To study the long-term variations of the magnetic fluxes in polar regions, we calculate the positive, negative, total, and net magnetic fluxes in the polar regions with latitudes above 70$\degr$. The entire surface area of the polar region above 70$\degr$ latitude is 1.84$\times$10$^{21}$ cm$^{2}$. The magnetic flux variations in the entire north polar region are displayed in Figure \ref{fig_flux_variation}a.
At the solar maximum, the positive magnetic flux (indicated by the asterisk) and negative magnetic flux (indicated by the diamond) were nearly balanced with the values of $\pm$0.5$\times$10$^{22}$ Mx, and the total unsigned flux was 1.0$\times$10$^{22}$ Mx. During the declining phase of the solar cycle, the positive flux, which dominated the north polar region, began to increase gradually, while the negative flux remained unchanged generally. At the solar minimum in 2019, the positive flux reached a peak of 1.3$\times$10$^{22}$ Mx, and the total flux and net flux peaked at 1.8$\times$10$^{22}$ Mx and 0.8$\times$10$^{22}$ Mx, respectively.
For the south polar region, the magnetic flux variation is shown in Figure \ref{fig_flux_variation}b.
At the polarity reversal stage of the south polar magnetic fields in 2014, the positive and negative magnetic fluxes were $\pm$0.6$\times$10$^{22}$ Mx. Correspondingly, the total unsigned flux was 1.2$\times$10$^{22}$ Mx, which is the minimum value during the solar cycle. After the polarity reversal, the magnetic fields in the south polar region began to be dominated by the negative polarity. Although the negative magnetic flux kept growing until 2017, the positive flux remained stable at about 0.5$\times$10$^{22}$ Mx. In 2017, the polar magnetic fields reached their maximum strength with the maxima of the negative flux, total flux, and net flux of $-$1.4$\times$10$^{22}$ Mx, 1.9$\times$10$^{22}$ Mx, and $-$0.9$\times$10$^{22}$ Mx, respectively.

At different phases of the solar cycle, the proportions of the positive and negative magnetic fluxes in the total flux in polar regions are shown in Figure \ref{fig_flux_percent}. For the north polar region at the solar maximum, the positive flux (indicated by asterisks) and negative flux (indicated by diamonds) were generally balanced. After the polarity reversal, the proportion of positive magnetic flux compared with that of the negative flux was slightly higher until the end of 2015. Then the proportion of the positive flux increased significantly and reached 70.9\% in 2019, i.e., the ratio of the positive to negative flux was 2.4. The variations of the flux proportion versus the thresholds of the radial flux density in 2019 are shown in Figure \ref{fig_flux_percent}b. The proportions of the positive flux (the red curve) and negative flux (the blue curve) increase and decrease as the radial magnetic field strength increases, respectively. When the unsigned radial flux density exceeds 50 Mx cm$^{-2}$, the proportion of the positive flux in the total flux is 83.5\%, i.e., the ratio of the positive to negative flux is 5.1. When the threshold of the radial flux density is set to 100 Mx cm$^{-2}$, the proportion of the positive flux in the total flux is as high as 94.5\%, i.e., the ratio of the positive to negative flux is 17.3.

For the south polar region before the polarity reversal, the proportions of the positive flux in the total flux in 2012 and 2013 were 67.6\% and 64.8\%, respectively. While after the reversal of magnetic polarity in 2014, the magnetic fields became dominated by negative polarity. The proportion of the negative flux in the total flux increased sharply from 2015 and reached a peak of 72.3\% (i.e., the ratio of the negative to positive flux was 2.6) in 2017. The variations of the magnetic flux proportion with different thresholds of the radial flux density in 2017 are displayed in Figure \ref{fig_flux_percent}d. The positive (the red curve) and negative (the blue curve) flux proportions decrease and increase, respectively, with the increase of the threshold of radial flux density. When the unsigned radial magnetic flux density exceeds 50 Mx cm$^{-2}$ and 100 Mx cm$^{-2}$, the proportions of the negative flux in the total flux are 85.1\% and 95.3\%, i.e., the ratios of the negative to positive flux are 5.7 and 20.3, respectively.

As the primary parameter for characterizing the vector magnetic structure, the inclination angle $\gamma$ with respect to the local normal is examined as shown in Figure \ref{fig_inclination}, in which both the ``vertical'' and ``horizontal'' magnetic fields are used.
In order to make a comparison between the positive and negative magnetic structures, the inclination angles of the negative magnetic fields are defined as 180$\degr$$-$$\gamma$ in the figure. At the solar cycle maximum, the magnetic inclination angles of both the positive and negative magnetic fields in the north polar region are around 27$\degr$ (Figure \ref{fig_inclination}a). During the following declining phase and at the minimum, the inclination of the negative fields did not change significantly, except for some fluctuations. However, for the positive fields, the magnetic inclination angles gradually decreased after the polarity reversal and reached 24.6$\degr$ in 2019. In 2019, the variations of magnetic inclination angle with the thresholds of radial magnetic flux density in the north polar region are shown in Figure \ref{fig_inclination}b. For both the positive (represented by the red curve) and negative (indicated by the blue curve) fields, the magnetic inclination angles decrease with the increase of the threshold of radial magnetic flux density. The inclination angle of the positive fields above any given threshold of radial flux density is smaller than that of the negative fields. For the thresholds of 50 Mx cm$^{-2}$ and 100 Mx cm$^{-2}$, the inclination angles of the positive fields are as small as 15.8$\degr$ and 13.4$\degr$, respectively. While those for the negative fields are 18.2$\degr$ and 16.6$\degr$, respectively.

Before the polarity reversal in the south polar region, the magnetic fields with positive polarity (i.e., the dominant polarity) had an average inclination of 24.2$\degr$ in 2012 (Figure \ref{fig_inclination}c). At the same time, the average inclination angle of the negative magnetic fields was 26.5$\degr$, larger than that of the positive polarity. At the polarity reversal stage in 2014, the inclination angle of the positive fields increased to 26.6$\degr$, comparable to that of the negative fields. After the polarity reversal, the inclination angle of the positive fields remained around 26.5$\degr$ with some fluctuations. On the other hand, after the polarity reversal, the negative polarity became the dominant polarity in the south polar region. Subsequently, the average inclination angles of the negative fields decreased rapidly and reached the minimum value of 23.8$\degr$ in 2018. The variations of the inclination angles with the thresholds of unsigned radial flux density in 2018 are displayed in Figure \ref{fig_inclination}d. The inclination angles of both the positive and negative fields show a decreasing trend when the threshold of the radial flux density increases. Above any threshold, the inclination angle of the negative fields is smaller than that of the positive fields. For the thresholds of 50 Mx cm$^{-2}$ and 100 Mx cm$^{-2}$, the inclination angles of the negative fields are as small as 15.5$\degr$ and 12.9$\degr$, respectively. For the same thresholds, the inclination angles of the positive fields are 17.9$\degr$ and 14.9$\degr$, respectively.

\section{Conclusions and Discussion}

\begin{figure*}
\centering
\includegraphics
[width=\textwidth]{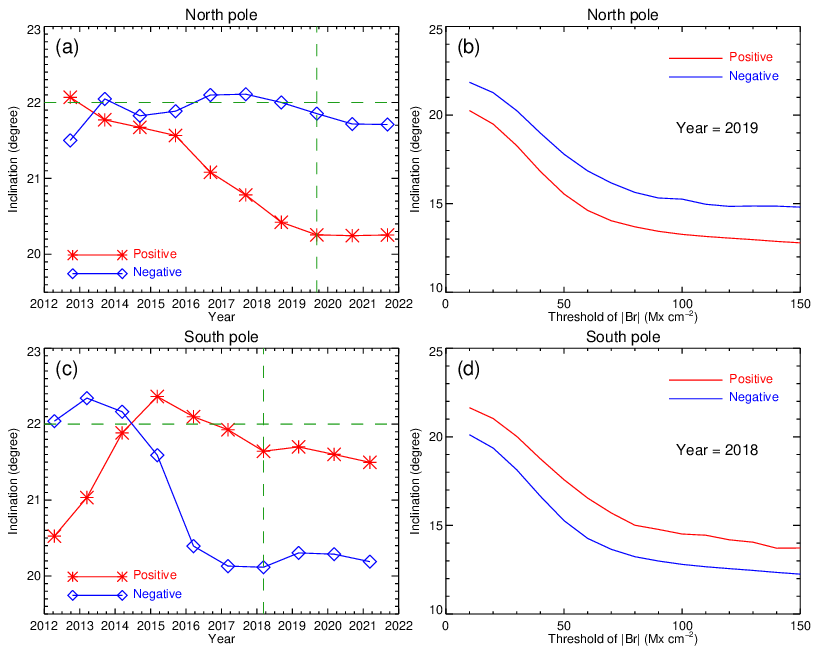} \caption{Similar to Figure \ref{fig_inclination} but only for the ``vertical" magnetic fields. \label{fig_inclination_v}}
\end{figure*}

\begin{figure*}
\centering
\includegraphics
[width=0.5\textwidth]{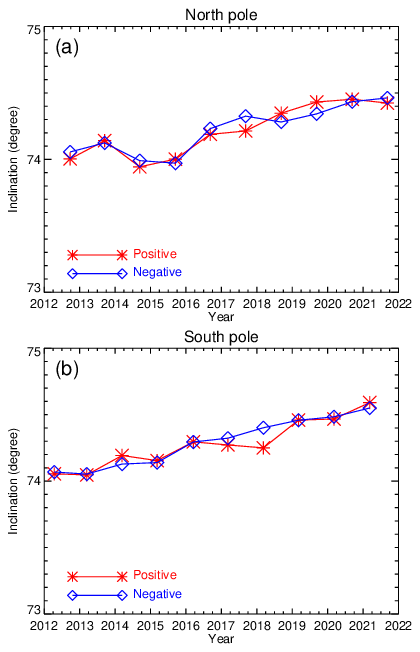} \caption{Long-term variations of the average inclination angles of the ``horizontal" magnetic fields. (\textbf{a}) and (\textbf{b}) Similar to Figures \ref{fig_inclination}(\textbf{a}) and \ref{fig_inclination}(\textbf{c}), respectively. \label{fig_inclination_h}}
\end{figure*}

With 10 years of polar vector magnetic field observations from Hinode/SP, we study the long-term variations of the magnetic flux, the flux proportion of different polarities, and the magnetic inclination with respect to the local normal in the polar regions above 70$\degr$ latitude during solar cycle 24.
In the north polar region, the positive flux increased from 0.5$\times$10$^{22}$ Mx in 2014 to 1.3$\times$10$^{22}$ Mx in 2019, while the negative flux remained generally unchanged at $-$0.5$\times$10$^{22}$ Mx.
In the south polar region, the negative flux increased from $-$0.6$\times$10$^{22}$ Mx in 2014 to $-$1.4$\times$10$^{22}$ Mx in 2017, while the positive flux remained stable at 0.5$\times$10$^{22}$ Mx.
After the polarity reversal, the proportion of the positive flux in the total flux in the north polar region increased to a maximum value of 70.9\% in 2019, and that of the negative flux in the south polar region peaked at 72.3\% in 2017. In the north polar region, the magnetic inclination of the positive fields varied from about 27$\degr$ at the solar maximum to a minimum value of 24.6$\degr$ in 2019, while that of the negative fields fluctuated around 27$\degr$. In the south polar region, the inclination of the negative fields decreased from 27$\degr$ to a minimum value of 23.8$\degr$ in 2018, while that of the positive fields remained at around 26.5$\degr$. For the magnetic fields in both the north and south polar regions, the inclination angles decrease with increasing threshold of radial magnetic flux density.

For the magnetic flux in the north polar region at the minimum of solar cycle 23, \cite{2010ApJ...719..131I} studied the polar data observed by Hinode/SP on 2007 September 25. In the field-of-view corresponding to a surface area of 8.4$\times$10$^{20}$ cm$^{2}$, the net magnetic flux was 1.7$\times$10$^{21}$ Mx. Thus the net flux for the whole north polar region above 70$\degr$ latitude was estimated to be 3.7$\times$10$^{21}$ Mx. As revealed in the present study (Figure \ref{fig_flux_variation}a), during solar cycle 24, the peak of the net magnetic flux in the north polar region was 0.8$\times$10$^{22}$ Mx in 2019, much larger than that in solar cycle 23.
As estimated by \cite{2008ApJ...688.1374T}, considering the nominal filling factor derived with the least-squares fit in the Milne-Eddington inversion, the total radial flux in the whole south polar region above 70$\degr$ latitude is 5.6$\times$10$^{21}$ Mx at the minimum of solar cycle 23.
As shown in Figure \ref{fig_flux_variation}b, we found that the total radial flux in the same area around the minimum of solar cycle 24 was 1.9$\times$10$^{22}$ Mx, also much larger than that in solar cycle 23.
In both polar regions after the polarity reversal, the magnetic fluxes of the non-dominant polarity remained stable at approximately 0.5$\times$10$^{22}$ Mx. This implies that a local dynamo, which populates in the lower latitudes \citep{2003ApJ...584.1107H,2014ApJ...781....7Y}, also exists in polar regions \citep{2009ApJ...706L.145S,2020ApJ...889L..26J}. According to in situ measurements from the Ulysses spacecraft, the magnetic flux above 36$\degr$ heliolatitude at 1 AU \citep{2000JGR...10510419M} is estimated to be approximately 2$\times$10$^{22}$ Mx \citep{2008ApJ...688.1374T}, which is somewhat larger than the polar photospheric magnetic flux obtained in this study.

The polar magnetic fields with opposite polarities were mainly balanced at the solar maximum, while they were significantly imbalanced at the solar minimum (Figure \ref{fig_flux_percent}).
For the flux proportion between opposite polarities at the minimum of solar cycle 23, \cite{2011ApJ...732....4J} focused on the north polar region observed by Hinode on 2007 September 10. They found that the ratio of the negative flux to the positive flux was 2, which means that most of the polar magnetic flux was closed. As shown in the present paper (Figure \ref{fig_flux_percent}), at the minimum of solar cycle 24, the proportion of the positive flux in the total flux reached 70.9\% in 2019, i.e., the ratio of the positive to the negative fluxes was 2.4, larger than that in solar cycle 23. In polar regions, there exist many closed loops connecting the opposite polarities, and the net magnetic flux is open \citep{2015LRSP...12....5P}. Considering the larger total magnetic flux and the higher imbalance between opposite polarities, we can deduce that more magnetic flux was opening into interplanetary space in solar cycle 24 than that in solar cycle 23. Since the polar field at a sunspot minimum is strongly correlated with the strength of the next cycle \citep{2007MNRAS.381.1527J}, our results can be used to explain why solar cycle 25 is stronger than solar cycle 24 as indicated by the sunspot number\footnote[2]{\url{https://sidc.be/SILSO/home}}.

At most time of the solar cycle except for the solar maximum, the magnetic inclination angles of the dominant polarity fields were smaller than those of the magnetic fields with non-dominant polarity (Figures \ref{fig_inclination}a, c). The reason may be that the non-dominant polarity fluxes exist as low-lying closed loops with large inclination, while most of the dominant polarity fluxes are almost radial open field lines with small inclination.
During the solar minimum, the magnetic inclination angles reached their minimum values, implying that the stronger the dominant polarity field, the more vertical the field lines.
With the increase of the threshold of radial magnetic flux density, the inclination angles of both two polarities decrease (Figures \ref{fig_inclination}b, d), revealing a fanning-out structure of the polar magnetic patches.
The results of \cite{2020A&A...644A..86P} also revealed that the polar magnetic field vectors within relatively large patches indicate a magnetic structure expanding with height. This kind of fanning out scenario was found to be a fundamental property of flux tube models \citep{1999A&A...347L..27S,2010A&A...518A..50P}.

The results shown in Figure \ref{fig_inclination} are derived from the data including ``vertical" and ``horizontal" fields. To compare intrinsic properties of two distinct populations, we separately investigate the inclination angles of the ``vertical" and ``horizontal" fields, as presented in Figures \ref{fig_inclination_v} and \ref{fig_inclination_h}.
Figure \ref{fig_inclination_v} shows the magnetic inclination angle using only the ``vertical'' magnetic fields. For the north polar region, the inclination of the positive vertical fields varied gradually from about 22$\degr$ at the solar maximum to a minimum value of 20.2$\degr$ at the solar minimum (Figure \ref{fig_inclination_v}a). While the inclination of the negative fields fluctuated around 22$\degr$ (marked by the horizontal line). The inclination variations of the positive (represented by the red curve) and negative (indicated by the blue curve) fields as a function of the radial flux density threshold are shown in Figure \ref{fig_inclination_v}b. For both polarities, the magnetic inclination angles decrease with increasing threshold of radial magnetic flux density, with a smaller inclination of the positive (dominant polarity) fields compared to that of the negative fields. For the threshold of 50 Mx cm$^{-2}$, the inclination angles of the positive and negative fields are 15.5$\degr$ and 17.8$\degr$, respectively. When the threshold is as high as 100 Mx cm$^{-2}$, the inclination angle of the positive fields is 13.3$\degr$ and that of the negative fields is 15.3$\degr$.
For the south polar region, both the positive and negative fields at the solar maximum had comparable inclination angles of approximately 22$\degr$ (Figure \ref{fig_inclination_v}c).
After the polarity reversal, the inclination of the positive fields remained around 22$\degr$, while that of the negative fields decreased to a minimum value of 20.1$\degr$ in 2018. The variation of inclination with flux density thresholds in the south polar region is displayed in Figure \ref{fig_inclination_v}d, which is similar to that in the north polar region shown in Figure \ref{fig_inclination_v}b.
At thresholds of 50 Mx cm$^{-2}$ and 100 Mx cm$^{-2}$, the inclination angles of the negative (dominant polarity) fields are as small as 15.3$\degr$ and 12.8$\degr$, respectively, while those of the positive fields are 17.6$\degr$ and 14.5$\degr$. These properties are very similar to those shown in Figure \ref{fig_inclination}.
We also examine the inclination angles of the ``horizontal" magnetic fields (Figure \ref{fig_inclination_h}). It is clear that, in both the north and south polar regions, the average inclination angles of the dominant polarity were almost the same as those of the non-dominant polarity. During the ten years, the average inclination angles of both polarities increased gradually from 74$\degr$ to 74.5$\degr$, and showed no clear correlation with the solar cycle. This trend may be due to the long-term throughput variation of SP \citep{2012ApJ...753..157S}. Therefore, the results shown in Figure \ref{fig_inclination} are mainly contributed by the ``vertical'' magnetic fields.

\begin{acknowledgments}
We are grateful to the referee for the constructive comments and valuable suggestions. This research is supported by the National Key R\&D Program of China (2022YFF0503800), Beijing Natural Science Foundation (1252034), the Strategic Priority Research Programs of the Chinese Academy of Sciences (XDB0560000), the National Natural Science Foundations of China (12350004, 12273061, 12273060, 12222306, 12573056, 12533010), the Youth Innovation Promotion Association CAS, and the Specialized Research Fund for State Key Laboratory of Solar Activity and Space Weather. Hinode is a Japanese mission developed and launched by ISAS/JAXA, with NAOJ as domestic partner and NASA and STFC (UK) as international partners. It is operated by these agencies in co-operation with ESA and NSC (Norway). ISEE Database for Hinode SOT Polar Magnetic Field (doi: 10.34515/DATA\_HSC-00001) was developed by the Hinode Science Center, Institute for Space-Earth Environmental Research (ISEE), Nagoya University.
\end{acknowledgments}



\end{document}